\documentclass[useAMS,usenatbib]{mn2e}
\usepackage{graphicx}
\usepackage{amsmath}
\usepackage{color}
\usepackage{setspace}

\DeclareMathOperator*{\erf}{erf}

\newcommand{\executeiffilenewer}[3]{%
 \ifnum\pdfstrcmp{\pdffilemoddate{#1}}%
 {\pdffilemoddate{#2}}>0%
 {\immediate\write18{#3}}\fi%
}

\title[Stellar Imaging with Cassini Occultations I]{High Angular Resolution Stellar Imaging with Occultations from the Cassini Spacecraft I: Observational Technique}
\author[Paul N. Stewart et al.]
{\parbox{\textwidth}{Paul N. Stewart$^{1}$\thanks{E-mail:p.stewart@physics.usyd.edu.au (PNS)},
Peter G. Tuthill$^{1}$,
Matthew M. Hedman$^{2}$,
Philip D. Nicholson$^{2}$,
and James P. Lloyd$^{2}$}\vspace{0.4cm}\\
$^{1}$Sydney Institute for Astronomy, School of Physics, H90, The University of Sydney, NSW 2006, Australia\\
$^{2}$Department of Astronomy, Cornell University, Ithaca, NY 14853, USA}
\begin{document}

\date{Accepted 2013 May 20. Received 2013 May 17; in original form 2013 January 22}

\pagerange{\pageref{firstpage}--\pageref{lastpage}} \pubyear{2013}

\maketitle
\label{firstpage}

\begin{abstract}

We present novel observations utilising the Cassini spacecraft to conduct an observing campaign for stellar astronomy from a vantage point in the outer solar system. By exploiting occultation events in which Mira passed behind the Saturnian ring plane as viewed by Cassini, parametric imaging data were recovered spanning the near-infrared. From this, spatial information at extremely high angular resolution was recovered enabling a study of the stellar atmospheric extension across a spectral bandpass spanning the 1 -- 5\,$\mu$m spectral region in the near-infrared.
The resulting measurements of the angular diameter of Mira were found to be consistent with existing observations of its variation in size with wavelength.
The present study illustrates the validity of the technique; more detailed exploration of the stellar physics obtained by this novel experiment will be the subject of forthcoming papers.
\end{abstract}


\section{Introduction}

Our limited ability to resolve fine angular detail in celestial targets presents an impediment to progress in many branches of astrophysics. For most existing instruments this limit is defined by the size of the aperture, and can be approximated by $\frac{\lambda}{D}$ where $D$ is the diameter of a telescope or $\frac{\lambda}{B}$ where $B$ is the longest baseline in an interferometer~\citep{Labeyrie2006}. For terrestrial observations in the optical and infrared, such theoretically optimum performance can be challenging to obtain due to distortion of the wavefront by atmospheric turbulence. In spite of significant technical progress in interferometry~\citep{Berger2012}, adaptive optics~\citep{Davies2012} and novel data analysis methods~\citep{Martinache2010}, the physical instrument scale imposes an inherent limit to most astronomical imaging.

Lunar occultations have been shown to provide high resolution observations using relatively small telescopes. For a star to be observed in this manner, it necessarily must be in the small region of the celestial sphere which is traversed by the path of the moon. This means that the majority of stars in the sky are out of reach of such a technique. Astronomical measurements using lunar occultations were first proposed in 1908~\citep{MacMahon1908}, and the use of diffraction for this purpose was described by~\citet{Eddington1909}. Having developed significantly over the last century the use of lunar occultations for scientific observations has a rich and fruitful history~\citep{White1987}, with major contributions by~\citet{Evans1953} and~\citet{Richichi1989}. These results include studies of binary systems~\citep{R94,R97,R00}, the stellar populations towards the galactic centre~\citep{R08,R11} and studies of the circumstellar environments of evolved stars~\citep{Ra99, T99, C01, M04}.

In this paper we report the first astrophysical measurements derived from stellar occultations by the rings of Saturn as observed by NASA's Cassini spacecraft. This technique provides very high angular resolution observations of bright stellar targets spanning the near infrared. These observations can yield an angular resolution as small as one milli-arcsecond (mas) with observations unaffected by the terrestrial atmosphere. To date, over 100 stellar occultation events have been observed by the spacecraft's VIMS instrument, targeting many different stellar objects. This paper analyses three of the earliest occultations of the asymptotic giant branch (AGB) star Mira in an exploration of the potential of the technique for astrophysical observations.
The selection of this star as a source was not initially made for the purposes of astrophysical investigation, but rather as a suitable source for Saturn ring science.
The decision was primarily based on its location in the sky and its brightness.
Mira is known to experience large amplitude variations with a period of 332 days~\citep{Ireland2011} and is of particular interest as the archetype of the Mira variable AGB star.
As a class undergoing heavy mass-loss, these stars contribute greatly to gas and dust enrichment of ISM and so are a key contributors to galactic chemical evolution.
Mira also has a compact companion~\citep{Ireland2007} which is not expected to be detected at our limits
and for which there was no evidence in our analysis.
Further astrophysical analysis of Mira and other targets will follow in subsequent publications.

\section{Experiment Description}
The VIMS instrument can be used to observe stellar occultations from the Saturnian system using various objects within the system as occulters~\citep{Brown2004}. These observations have been used in order to probe the atmospheres of Saturn and Titan and other moons, as well as the nature of the ring system~\citep{Nicholson2005, Shemansky2005, Hedman2007, Colwell2010, Nicholson2010, Hansen2011, Nicholson2011}. Some edges within the rings are extremely compact, exhibiting dramatic changes in opacity on the order of ten metres radially~\citep{Colwell2010}. Such edges can be used as sharp-edged, semi-infinite plane occulters, providing an opportunity to measure the starlight in the Fresnel diffraction pattern propagated from this edge to the spacecraft.

These observations are in principle analogous to lunar occultations with several useful advantages.
Cassini's distance from the RPI was approximately four times the Earth-Moon distance at the time of the observations.
This gives the diffraction pattern more distance over which to spread and become easier to resolve when compared to terrestrial lunar occultations.
This distance varies with each occultation, and for these particular events was especially distant. Observations were uncorrupted by scintillation noise inherent to ground-based data. The rings at this time covered 9 degrees of the sky from Cassini's position, enabling access to many more potential targets than the half degree disc of the moon.
Finally the highly inclined orbit of the spacecraft allows for vastly greater numbers of potential occultations than the nearly equatorial orbit of the moon.
There are existing Cassini occultation observations of targets with ring opening angles ranging from 74$^\circ$ to -61$^\circ$, and as shallow as 2.4$^\circ$.

\subsection{The VIMS Instrument}
The Cassini Visual and Infrared Mapping Spectrometer (VIMS) is an imaging spectrometer, sensitive to a range of wavelengths from 0.3-5.1$\,\mu m$. It was constructed to observe both scattered and emitted light from surfaces and atmospheres within the Saturnian system with an emphasis on the spectral domain~\citep{Brown2004}. The instrument consists of two parallel telescopes feeding separate visible and infrared spectrograph channels. The infrared side, as used for the observations in this paper, samples from 0.8\,$\mu m$ to 5.1\,$\mu m$ at a resolution of 16.6\,$nm$~\citep{McCord2004}. VIMS was specifically designed to observe both stellar and solar occultations, and is the only instrument on board capable of such observations in the visible and infrared.
In occultation mode VIMS observes a single pixel, 0.25$\times$0.50 $mrad$ in size which is spectrally dispersed onto a linear detector giving simultaneous measurements in 256 bands.
It is capable of observing at speeds of up to 77Hz across its entire spectrum.
The observations used in this paper were binned down to 31 spectral channels and limited to 80\,ms exposures in order to conserve downlink time.
They were intended for ring rather than stellar studies and did not require faster exposures or finer spectral resolutions.
Cassini acquires a lock on the star using both solid state gyros and a star-tracking CCD, and once a lock is obtained the pointing has been found to be stable for many hours to the order of 0.05 mrad.
A detailed technical description of the VIMS instrument and its observing modes can be found in~\citet{Brown2004}.

\subsection{General Considerations}

\begin{figure} 
\centering
  \def\svgwidth{\columnwidth}
 \executeiffilenewer{pov.svg}{pov.pdf}%
 {inkscape -z -D --file=pov.svg %
 --export-pdf=pov.pdf --export-latex}%
 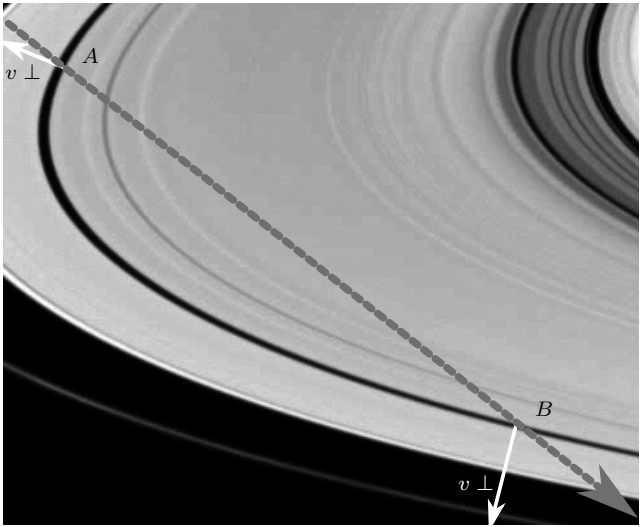%

  \caption{Schematic illustration of the path of the RPI across the rings in the sky-plane of Cassini. The dashed line represents this apparent path or the star behind the rings, with constant temporal and angular sampling. Points A and B show occulting edges within the rings. The white vectors represent $v_{\perp}$ for each occultion event. Background Image from \citet{CIT2010})}
  \label{fig:pov} 
\end{figure}

Starlight emitted from the distant stellar system is diffracted by sharp edges within the rings. The spacecraft is not travelling in the plane perpendicular to the line of sight to the star, so changes to the diffraction pattern with distance must be considered, giving a trajectory through a three dimensional diffraction field. The speed of the spacecraft through this field at a fixed sampling rate determines the spatial resolution of the observations. A slower perpendicular velocity ($v_{\perp}$) relative to the ring edges in the sky plane gives a higher angular resolution -- this is equivalently obtainable with a faster read out rate but fundamental limitations with photon noise (explored later) means this is only possible with brighter objects. Figure~\ref{fig:pov} shows the apparent path of a star behind the rings as viewed from the point of view of Cassini. In this frame the angular separation between samples is equal and is determined by the orbital velocity of the spacecraft and the integration time.

The spatial sampling of the diffraction field is in the direction normal to projection of the ring edges in Cassini's sky plane. As a consequence the spatial resolution obtained at the edge marked `A' is much coarser than at the edge labelled `B'. This normal to the diffracting edge also determines the direction in which the stellar image which is being sampled, allowing the recording of brightness profiles in different directions across the stellar surface. `A' and `B' in Figure~\ref{fig:pov} demonstrate an event geometry with sampling in approximately orthogonal directions as illustrated by the white lines marked $v_{\perp}$. In future work we will attempt to combine these different slices using tomographic techniques to produce high resolution 2D images of asymmetric targets.

Ground based lunar occultations have a mean fringe velocity of 0.6\,m/ms~\citep{Richichi1989}. This is the equivalent parameter to our $v_{\perp}$.
The occultations observed from Cassini have perpendicular velocities in the range 0.5 - 2.2\,m/ms which do not vary faster than 0.6\,m/s/s.
If the distance from Cassini to the RPI was reduced to be consistent with lunar occultations, this range reduces to 0.12 - 0.53\,m/ms.
Whilst these values are smaller than those available for ground based lunar occultations, it is important to note that VIMS is limited to minimum 13\,ms integrations.
Terrestrial cameras are capable of running much faster, with occultations being recorded at as fast as 3.2\,ms~\citep{Richichi2012}.

\subsection{First Observations}

In mid 2005 the Saturnian rings occulted Omicron Ceti, or Mira, on three subsequent orbits. These have been designated Rev 8, 9, and 10. The star was observed well prior to its encounter with the outer edge of the F ring and tracked as it passed behind the various features of the ring system. 
Starlight was visible through the rings within the Encke Gap, and was occulted by its hard edges as the spacecraft moved along its trajectory.
Results from these occultations were first published by \citet{Hedman2007}, and revealed fine scale texture in the A ring, and large variations in transmission with longitude of the occultation track relative to the direction to the star.
These observations also form the basis of this paper, using the sharp inner edge of the Encke Gap as a semi-infinite opaque occulter, with the aim to observe structure in the source star. 

This particular edge was used exclusively for this study as an example of one of the sharpest and cleanest edges in the ring system, and to provide consistency between model fits.
The mean normal optical depth of the A ring is around 0.5, but due to the very low inclination of the rings which occurs when observing Mira (-3.45$^\circ$), it becomes effectively opaque.
The constituent particles of the ring range from centimeters to meters in size, and are much smaller than the Fresnel zone of ~100\,m.
The resulting optical wavefront will therefore average over many individual particles.
The A ring has been shown to be less than 10\,m in thickness~\citep{Colwell2010, Jerousek2011}.
Edges like these have been shown to have radial widths of the same order.

The Encke gap is kept clear, and its sharp edges maintained, by the shepherd moon Pan.
The motion of this satellite also causes the formation of small amplitude, long wavelength, radial perturbations on the edges of the gap.
These waves have a wavelength on the order of 1000\,km and an amplitude of up to 1\,km.
The change in the slope of the occulting edge can change by up to half a degree which will affect the projected perpendicular velocity of the RPI.
For the occultation on the egress of Rev 10, the shallowest angle and most affected edge of those in this paper, this causes an uncertainty of 4\% to the projected perpendicular velocity which follows though to the fitted radii.

\subsection{Dynamic Orbital Geometry}

\begin{figure} 
\centering
  \def\svgwidth{\columnwidth}
 \executeiffilenewer{geom_reduced.svg}{geom_reduced.pdf}%
 {inkscape -z -D --file=geom_reduced.svg %
 --export-pdf=geom_reduced.pdf --export-latex}%
 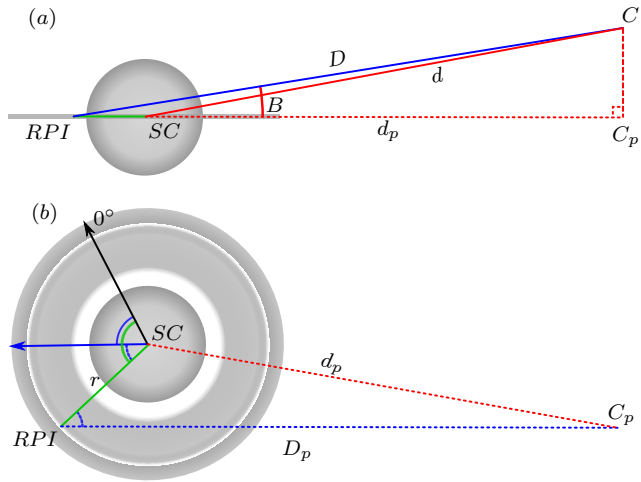%

  \caption{Views of the geometry of a single temporal sample within an occultation event depicted from an equatorial (a), and a polar (b) perspective. The centre of Saturn is labelled $SC$, the position of the Cassini spacecraft is labelled $C$, and the point on the line of sight from Cassini to the star which passes the ring-plane is labelled $RPI$. $D$ is the critical distance between Cassini and the $RPI$, and $d$ is the distance from the centre of Saturn to Cassini. Labels with a subscript ``$_p$'' are projections into the equatorial plane, which is also the plane of the rings. The angle $B$ is the opening angle of the rings to the Cassini-star line of sight. The blue vector in (b) is the line of sight to the star from Saturn's centre and is parallel to $D_p$. The vector marked $0^{\circ}$ is the zero longitude of the Saturnicentric coordinates and is defined by the node of the ring plane on the terrestrial equatorial plane (J2000 RA 130.589$^\circ$ Dec 0$^\circ$). The values of $r$ and 
longitude for the RPI, and the Cartesian coordinates of Cassini, are both included with the spacecraft ephemeris in the data headers.}
  \label{fig:geom_reduced} 
\end{figure}

An occultation event occurs during the precise line of sight alignment of three bodies. The geometry of the event is therefore critical in understanding the lightcurve observed by VIMS. In these observations it is assumed that the star and planetary rings are fixed, the instrument is looking directly at the star throughout the event, and as the spacecraft moves the edges within the rings obscure the line of sight to the star. To avoid any ambiguity, the point on the line of sight from Cassini to the star at which the ring plane is intercepted will be referred to as the Ring Plane Intercept (RPI) henceforth. Key information to reconstruct the event is available via SPICE Kernels (from the NASA Planetary Data System's Navigation Node), including the position of the spacecraft in Cartesian coordinates. From this and the stellar coordinates we can calculate the latitude and the radius within the ring plane of the RPI. 
This provides the positions of both the spacecraft and the RPI in a Saturnicentric reference frame as is detailed in Figure~\ref{fig:geom_reduced}.

The distance from Cassini to the RPI is critical for the determination of diffraction effects and can be calculated from the positional information mentioned above. This distance is calculated by taking the vector difference between the Saturnicentric positions of the spacecraft and the RPI. This is the distance over which the starlight is diffracted around the edge of the rings before being recorded by Cassini. The relative velocity of the spacecraft is determined from the recorded change in position. With this distance determined, and the position of the RPI projected into Cassini's sky-plane as shown in Figure~\ref{fig:pov}, we have everything required to determine the point source light curve of any observed occultation event.

\subsection{Achievable Results}

The obtainable angular resolution limit of this technique is primarily dependant on the geometry of the event.
Most significantly by the distance from Cassini to the RPI and the combination of instrument sampling speed and radial velocity of the RPI, which gives angular sampling.
Such angular sampling would ideally be less than the angular separation between the first maxima and first minima for a point source lightcurve.
This is the Fresnel diffraction fringe spacing ($\theta_F$) and is proportional to the radius of the first Fresnel zone.
In practice it is still possible to fit models to slightly undersampled data, especially for a large, resolved star like Mira.
In Table~\ref{tab:obs} the specifics of the occultation events are shown.
The finest sampling resolution ($\theta_S$), 4.69 $mas$, is provided by the egress of Rev. 10, which is insufficient to fully sample of the largest fringe spacing.
This could have been improved if the integration time was 20 $ms$ which would have yielded a sampling resolution of just over 1 $mas$. With still slower perpendicular velocities and comparable or longer Cassini-RPI distances it is in principle possible to obtain sub-milli-arc-second sampling resolutions. 

\begin{table}
\centering
\caption{Table of observations. The date is the day of the year in 2005, $\phi$ is the approximate phase of the stellar pulsation, $D$ is the distance from Cassini to the RPI, $v_{\perp}$ is the component of the velocity of the RPI perpendicular to the rings in the sky plane, $\theta_S$ is the sampling resolution, and $\theta_F$ is the Fresnel diffraction fringe spacing at 3\,$\mu m$. The events are labelled by the orbit number and annotated for ingress (i) or egress (e).}
\begin{tabular}{c|c|c|c|c|c|c|c}
Event&Date & $\phi$ & $D$ & $v_{\perp}$  & $\theta_S$  & $\theta_F$ \\
& (DOY) &  & ($10^6 km$) & ($km/s$) & ($mas$) & ($mas$) \\ \hline

8i&144 & 0.2 & 1.568 & 1.125 & 12.1 & 4.12 \\
8e&144 & 0.2 & 1.736 & 0.709 & 6.90 & 3.92 \\
9i&162 & 0.25& 1.574 & 1.378 & 14.8 & 4.12 \\
9e&162 & 0.25& 1.684 & 0.634 & 6.37 & 3.98 \\
10i&180 & 0.3& 1.595 & 2.082 & 22.1 & 4.09 \\
10e&180 & 0.3& 1.643 & 0.455 & 4.69 & 4.03 \\
\end{tabular}
\label{tab:obs}
\end{table}

\section{Data Reduction and Model Fitting}

\begin{figure}
\includegraphics[width=.99\columnwidth]{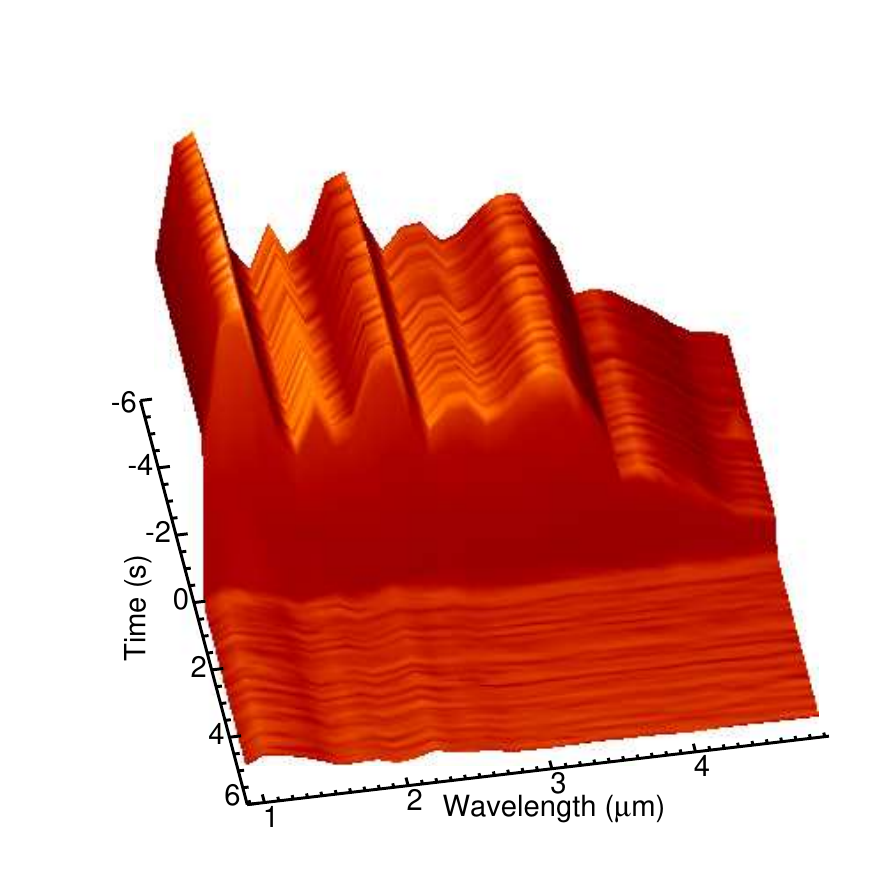}
\caption{Spectrally complete data obtained from an occultation by the inner edge of the Encke Gap on the ingress of Rev. 9. Spectral features of the unocculted starlight can be seen along the wavelength axis. The temporal axis which runs up and down the page shows the passage of time relative to the RPI crossing the occulting edge. The (unlabelled) vertical axis is intensity in counts as recorded by VIMS. The spatial information obtained using occultation methods is all found in the functional form of the ``cliff edge'' around zero seconds (the point of occultation) for each wavelength band.}
\label{fig:raw1}
\end{figure}

Each observation contains data covering 31 spectral channels divided between 0.94 and 4.93\,$\mu m$, and is selected to span several seconds either side of the occultation event. The 80\,$ms$ temporal sampling from the spacecraft moving through the radial direction of the diffraction field yields spatial sampling of the diffraction pattern. During each orbit, the star passes the inner edge of the Encke Gap twice, once on ingress and again on egress. Figure~\ref{fig:raw1} shows the raw data acquired on the egress during Cassini orbit 9. The geometrical occultation event itself is labelled $0\,s$. The region prior to this shows the spectra of the unocculted star. After occultation the spectrum of ringshine is dominant, but there may be a small amount of obscured starlight.
This is reflected sunlight from the rings entering the relatively large angular aperture of VIMS.
Note that although ringshine contributes the background flux for most events, there are a significant minority of occultations which occur when the RPI is in the shadow of the planet that are therefore devoid of ringshine. 

With a relatively long data sequence recorded both before and after the event, the stellar flux and the background levels (ringshine plus residual transmission and instrument thermal background) can be accurately calibrated over all spectral channels. High angular resolution information about the celestial target is then encoded by the detailed way in which the intensity varies around $t=0$. Because VIMS is a spectrometer, we may explore the spatial extension of the target star simultaneously across the near infrared using a single event. 

\subsection{Data Preparation}

The first step in data processing is the removal of cosmic rays and the subtraction of backgrounds. Cosmic rays are identified when a single temporal sample has some spectral data which are intensity outliers far beyond the normal range of the noise. These spurious data points (which are uncommon) are simply flagged and excluded from further analysis and model fitting. 
The ability to handle non-uniform time sampling is also a requirement in order to process data with regular pauses, or with any other temporal variation in data collection.
All samples are unambiguously tagged in time.
The occultations in this paper are rapid with the the slowest spanning only two seconds.
During this event $v_{\perp}$ changes by 0.15\% which has a negligible effect allows any potential effect of a changing occultation speed to be omitted.

The mean intensity of the data acquired when the star is behind the ring for each observation is subtracted from the entire light curve, removing the effects of ringshine, partial stellar transmission through the rings and any instrumental background. This process shifts the zero point of the intensity curve. For computational ease in later model-fitting steps the data are then normalised by the mean of the unocculted starlight for each band. This yields consistent lightcurves ranging from around zero to one in all spectral channels, with spectral intensity information being preserved separately.

The spatial separation between samples is determined by the distance travelled by the spacecraft in this time perpendicular to the ring edge. Most samples are separated by the operating sampling rate of the instrument, which for these observations includes 80\,$ms$ integration time and ~2\,$ms$ instrumental reset time. However every 64 samples, VIMS takes background data (which is saved separately) resulting in episodic larger gaps in the data record. Further missing samples may be produced by the flagging of cosmic rays. Our analysis software therefore makes no assumption about even sampling in the time domain. The projected displacement of the RPI from the geometric edge for each sample is calculated and recorded for use as an independent variable denoted `$x$' in the Figures below.
Finally the standard deviation in the areas of the lightcurve away from the occultation event is recorded and used as a measure of the characteristic noise levels in the data which contributes to uncertainty in the model fitting process.

\subsection{Lightcurve simulation}

As the occultation event proceeds, the intensity recorded as a function of time is referred to as the {\it lightcurve}. The diffraction of light by a semi-infinite plane generates a lightcurve which is the well known knife-edge diffraction pattern. For our observations, the sharp edges within the rings can be assumed to be a hard-edged semi-infinite plane, obscuring starlight reaching the spacecraft. The equation for the functional form of the amplitude of the spatial diffraction pattern for a point source is given by:

\begin{equation}\label{eq:fresnel}
    A=\frac{1}{\sqrt{2}}\left[\left(\mathcal{C}(\nu)+\frac{1}{2}\right)+i\left(\mathcal{S}(\nu)+\frac{1}{2}\right)\right] 
\end{equation}

where $\nu=x\sqrt{\frac{2}{\lambda D}}$, $\mathcal{C}(\nu)$, $\mathcal{S}(\nu)$ are the cosine and sine integrals, $x$ is the spatial ordinate orthogonal to the occulting edge and is set by the instrument's sampling rate and the spacecraft's relative velocity. The observed intensity is defined by $I=A^*A$. Evaluation of these Fresnel integrals is computationally difficult, but fortunately they can be extended into the complex plane and expressed in terms of error functions:

\begin{equation}\label{eq:Fresnel_ints_erfc}
  \mathcal{C}(\nu)=\frac{1-i}{4}\left[\erf\left(\frac{1+i}{2}\sqrt{\pi}\nu\right)+i\:\erf\left(\frac{1-i}{2}\sqrt{\pi}\nu\right)\right]
\end{equation}

\begin{equation}\label{eq:Fresnel_ints_erfs}
  \mathcal{S}(\nu)=\frac{1+i}{4}\left[\erf\left(\frac{1+i}{2}\sqrt{\pi}\nu\right)-i\:\erf\left(\frac{1-i}{2}\sqrt{\pi}\nu\right)\right]
\end{equation}

In this form the cosine and sine integrals are directly computable from functions in standard numerical libraries and can be used to build a lightcurve for any specific event geometry. The simulation of these lightcurves, along with the software discussed over the remainder of this paper, was performed with IDL code written especially for these datasets.

The lightcurve from any extended object can be modelled as the incoherent superposition of point sources at all positions on its surface. This can be calculated by convolving the point source lightcurve above with a function describing the source geometry. The two dimensional source function in the plane of the sky, when observed with occultation techniques, delivers a one dimensional projection of the target normal to the occultation edge. Thus our data are well suited to exploring intensity profiles summed over all points parallel to the occulter. This needs to be considered when building models for the interpretation of lightcurves in an astrophysical context. Although for an arbitrary extended source it is not possible to recover full two dimensional information with a single occultation event, in practice many objects may justify the assumption of circular symmetry thereby allowing the construction of meaningful models.

\subsection{Model Fitting}

Model fitting is a powerful way of extracting unknown parameters from recorded data. The lightcurves were compared to those generated from simple stellar models in order to determine the angular size of the star.
This was performed using the bisection method.
There has also been significant work done in fitting stellar sources more effectively than with single parameter fits. These include multivariate model fitting and model independent analysis.
Both of these techniques reduce the residuals substantially and will be detailed in future papers along with the results they provide.

For particularly fast occultations a point source lightcurve becomes a simple step function.
In this {\it geometric optics} regime the effects of diffraction can be neglected.
Consequently a model independent brightness profile can be recovered simply by differentiating the observed lightcurve.
Using such a geometric optics approximation for our VIMS data was found to be valid for the fastest events, but caused errors of up to 6\:$\sigma$ on model fits for slow well-sampled events.
Because most events fell into a regime where diffraction needed to be considered, the same code was applied uniformly to all data modelled here.

\subsubsection{Fitting for Stellar Radius}

\begin{figure}
\includegraphics[width=.99\columnwidth]{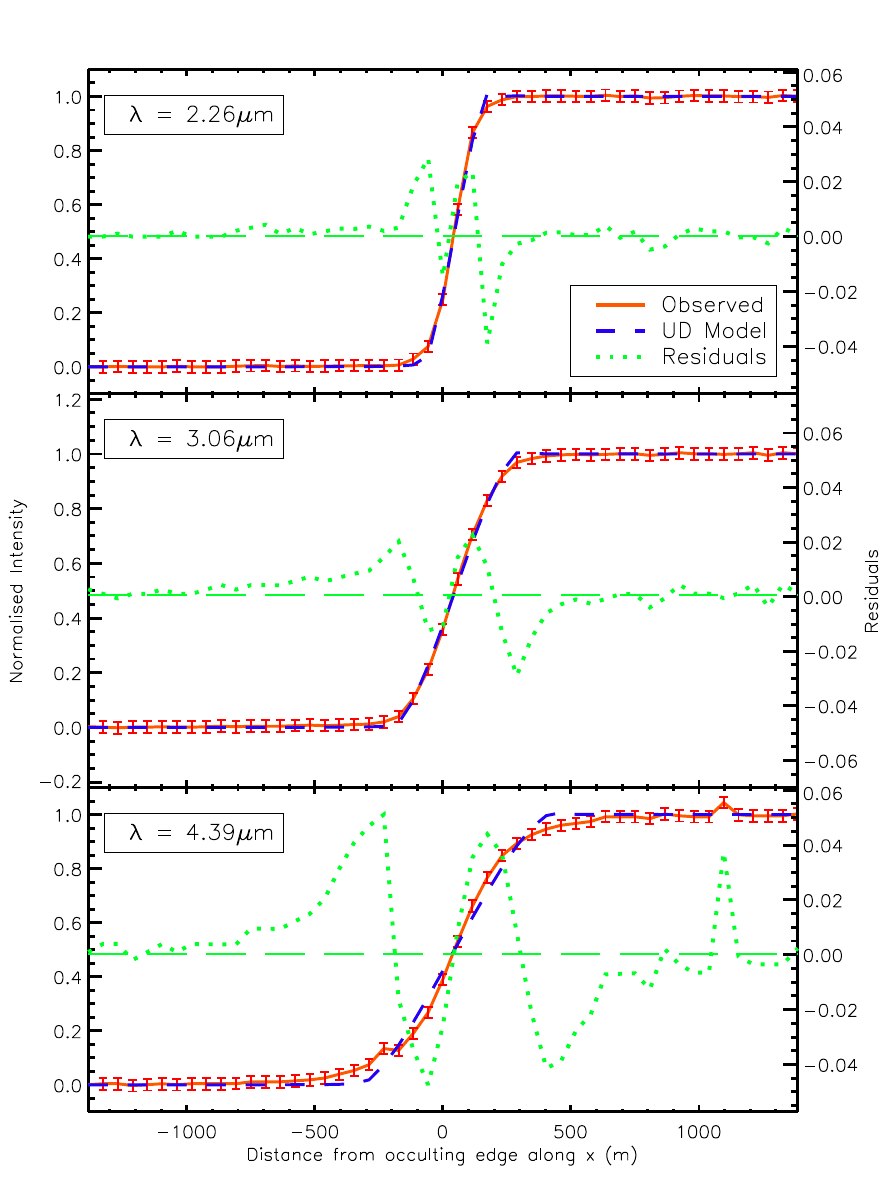}
\caption{The Observed lightcurve in three bands (2.26, 3.06 and 4.39\,$\mu m$) for the Encke Gap inner edge occultation on the egress of orbit 8 with $1 \sigma$ error bars (orange solid). A fitted Uniform Disc model (blue dashed) is also plotted against the normalised left hand scale. The residuals of the fit are plotted in green (dotted) with an expanded scale in the same units on the right. The horizontal axis shows the position of the RPI from the ring edge as projected into the sky-plane of Cassini.}
\label{fig:UD}
\end{figure}

A simple uniform circular disc model with a single free parameter -- the stellar radius -- provides a simple and robust measurement of overall size while also offering direct comparison with a wide variety of literature estimates. The shape of the $\chi^2$ space in the vicinity of the likely radius was found to have an unambiguous single minimum. The aim of model fitting is to find the radius of the model which gives this minimum. Initial limits were set both significantly above and below the expected result, allowing the minima to be identified with high accuracy in few iterations using a search based on the bisection method of root finding. This provided significantly higher accuracy and much greater efficiency than a grid search.

Figure~\ref{fig:UD} shows typical examples of fitting a uniform disc model to lightcurves at three different wavelengths in one observation.
These three bands are representative of the changes in angular size observed.
These bands are 2.26\,$\mu m$ which is amongst the smallest observed angular sizes, 3.06\,$\mu m$ which shows a moderate amount of circumstellar extension and 4.39\,$\mu m$ which corresponds to the largest observed size and the most extension.
For the most part, the fits are very close to the shape of the observed lightcurve, although there are some significant residuals remaining in the vicinity of the geometric edge, especially for wavelengths which correspond to larger angular sizes.
This suggests that a uniform disc provides a reasonable first-order fit, but that it is not capable of reproducing much of the fine structure contained in the data. In particular the shape of the lightcurve in the most critical regions at star's edges are not accurately modelled, with the peaks of the residuals being an order of magnitude higher than the RMS noise levels of $2.1\times 10^{-3}$, $3.2\times 10^{-3}$, and $7.9\times 10^{-3}$ in order of increasing wavelength. The worst excursions in the residuals occur at the leading and trailing edges of the disc, suggesting that the hard edged uniform disc model is a relatively poor representation, and a softer edge, either through limb darkening, the existence of circumstellar structure, or a combination of these may fit better.
Despite its shortcomings, a uniform disc fit gives a useful and uncomplicated gauge of the overall apparent size of the star, and is furthermore very important in comparing our results with literature values in which it is almost universally used.

\subsubsection{Polychromatic Fitting}

\begin{figure}
\includegraphics[width=.99\columnwidth]{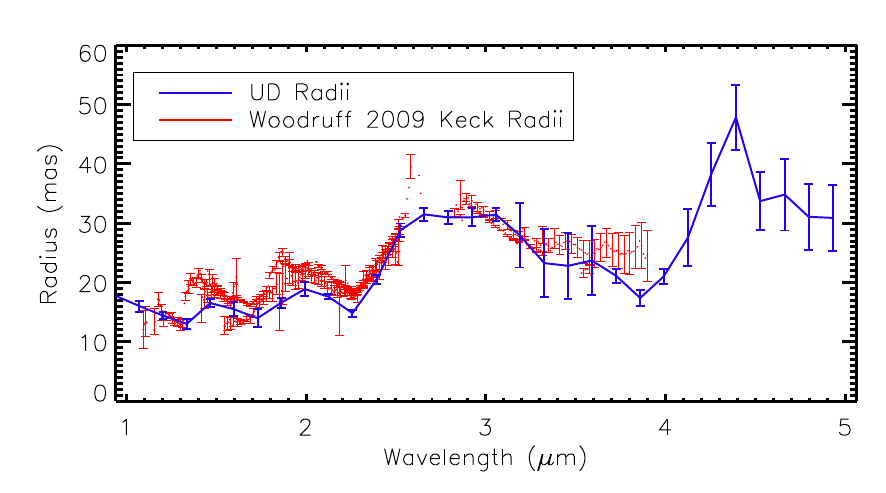}
\caption{The fitted uniform disc radii across all observed wavelengths for a single occultation event. The fitted uniform disc radii are shown in blue with one sigma error bars. The red data shows previously published interferometric measurements from \citet{Woodruff2009}.}
\label{fig:UD_all}
\end{figure}

The angular size of the star at all observed wavelengths was measured by fitting a uniform disc model to each recorded band as shown in Figure~\ref{fig:UD_all}.
As noted above, the observed angular size of the star varies substantially with wavelength, meaning that the star appears to be different sizes depending on the observed colour.
The three wavelengths shown in Figure~\ref{fig:UD} were found to have best fitting radii at 14.8$\pm$0.6, 31.5$\pm$1.1, and 47$\pm$7\,mas in order of increasing wavelength.
The errors were generally larger for both longer wavelengths and those with larger radii, but there were exceptions to this in the region between 2.5 and 3.0\,$\mu m$ where they remained relatively small.

This variation in stellar size is due to absorption and emission from extended molecular layers (H$_{2}$O, CO$_{2}$) within the stellar atmosphere~\citep{Woodruff2009, Yamamura1999} and will be discussed in more detail in a subsequent publication.
Figure~\ref{fig:UD_all} also shows that the fitted uniform disc radii are consistent with those previously published by~\citet{Woodruff2009}, plotted in red. The consistently slightly lower radius can be explained by the inherent variable nature of Mira type stars. Further discussion of the astrophysical interpretation of these results based on the wider dataset from~\citet{Stewart2011}, is beyond the scope of the present paper and will appear in a forthcoming publication.

\section{Noise and Limitations}

\begin{figure}
\includegraphics[width=.99\columnwidth]{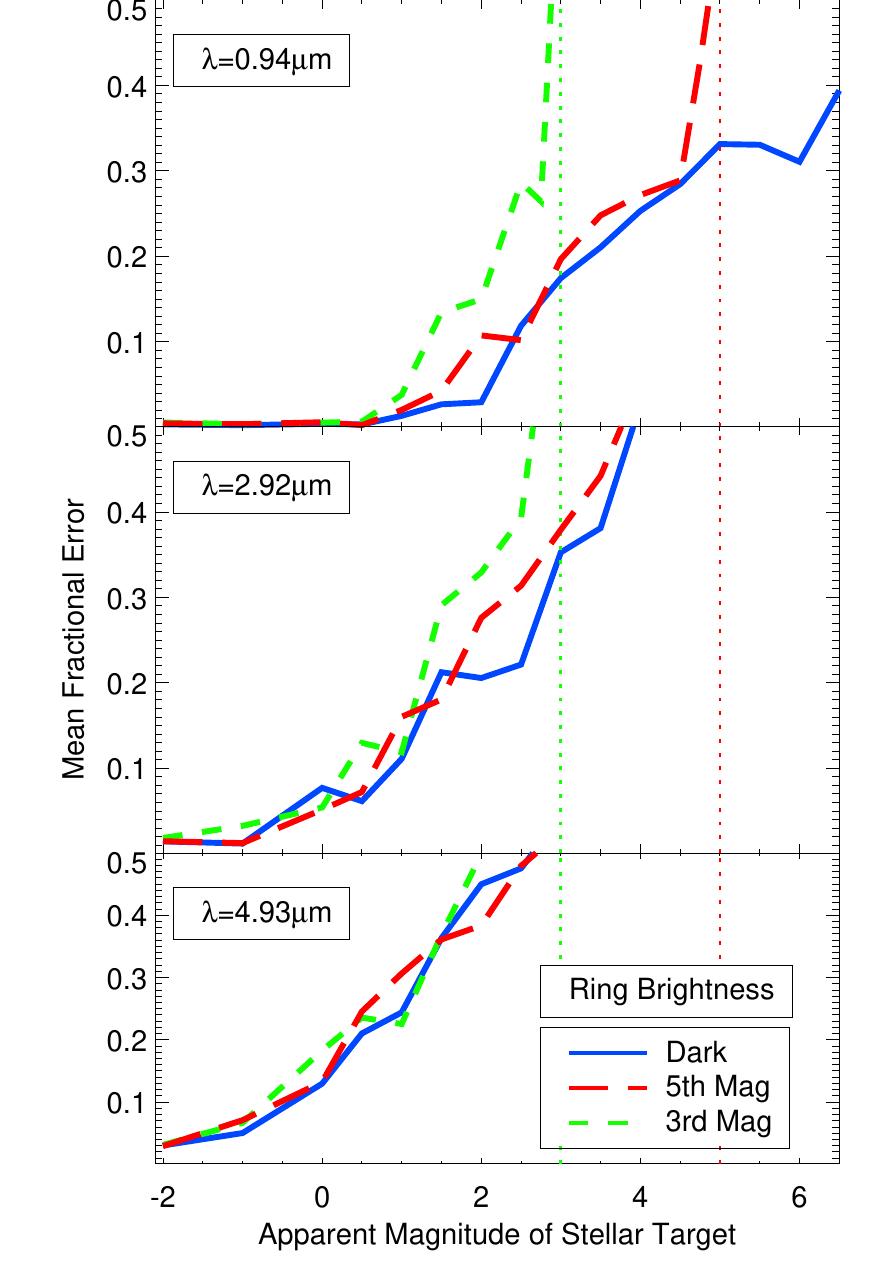}
\caption{These plots show the mean fractional fitting error simulated 20\,mas radius Uniform Disc models with 20\,ms integration times with varying stellar magnitudes and ring brightness. Each data point is the mean of one hundred fits to simulated data with unique artificial instrumentation noise. The three sub-plots show results for three wavelengths across the spectral range of VIMS. The three different lines in each plot compare results for different ring brightness ranging from third magnitude down to an undetectably dark ring as detailed by the legend. The vertical dotted lines are the cut-off points where a stellar target would have the same brightness as the occulting ring edge and correspond to the curves of the same colour. The spectra of both the star and ringshine are assumed to be (somewhat unrealistically) uniformly bright at all wavelengths. The magnitude of the ringshine is taken to be the sum of all flux falling within one temporal sample.
All magnitudes are in the Johnson UBVRI system.}
\label{fig:noise}
\end{figure}

The limitations of the technique were examined through the creation of an artificial lightcurve simulator. This was based on the noise characteristics of the VIMS instrument as previously detailed by~\citet{Miller1996} and were used from first principles to build a realistic model of instrumental noise. This allowed artificial lightcurves to be generated for model objects of idealised and known geometry. These lightcurves were fed into the same analysis software as the observations and compared in order to characterise the limiting parameters of such observations for astronomical purposes.

There were several parameters which affected the quality of the observables. The most important were the integration time of the instrument, the bandwidth of the observations, the magnitude of the star, and the brightness of the occulting ring edge. The quality of the model fitting was found to be independent of the angular size of the uniform disc model star for resolvable stars.

Figure~\ref{fig:noise} shows how fractional error is primarily influenced by the brightness of the stellar target, and to a lesser extent, by the brightness of the occulting ring edge. It can be seen that for longer wavelengths, ring brightness plays little part in the goodness of fit, whilst for shorter wavelengths the change in ring brightness is significant. It is also clear that shorter wavelengths provide superior fits to longer wavelengths, with errors below 10\% extending to stars approximately two magnitudes fainter. Ringshine, the brightness of the rings, varies significantly with radius and opening angle to the sun.
It also changes with the phase angle between the sun, the rings and the observer. As the brightness of the star approached that of the rings, the fractional error increased asymptotically, whilst with completely dark rings it was possible to get good fits for targets down to 5th magnitude at 0.94\,$\mu m$. For 4.93\,$\mu m$ the fits were only reasonable down to 1st magnitude.

Integration time was tested over the range from 13\,ms up to 80\,ms with limiting stellar magnitudes ranging respectively from 2nd down to 4th magnitude for 2.2\,$\mu m$ observations. This explored the possibility that objects significantly dimmer than those previously observed may make potential targets for future observations. Decreasing the integration time reduces the angular separation between samples, which improves the resolving power of the technique, but this can can only be done for brighter targets.

Co-adding of contiguous sets of spectral channels within VIMS allows the sensitivity to be increased at the expense of spectral resolution. Although care needs to be taken for objects with complex simultaneous spatial and spectral structure (such as spectral lines which arise in spatially distinct regions), in general sensitivity limits for grey or approximately grey objects might be dramatically improved by such binning (in addition to incurring lower data volumes for transmission from Cassini back to Earth).

\section{Conclusions}

We have demonstrated the scientific utility of occultations observed from the Cassini spacecraft in which targets of astrophysical interest pass behind the Saturnian ring plane. Such observations have many immediate advantages over the well established lunar occultation technique including the absence of scintillation noise, the opening of spectral windows unavailable from Earth, more favourable geometry for the events, and greater sky coverage for potential science targets.
Unlike interferometry, occultations also allow simultaneous sampling of a wide range of spatial scales.
We are also able to probe inside the significant water bands which are obscured by the Earth's atmosphere for any terrestrial observations.

The angular resolution provided by occultations at the highest VIMS cadence (13\,$\mu s$) is in principle very high, and for a sufficiently bright target would be competitive with an optical interferometer with kilometric baselines.

The technique could even be extended into the hunt for exoplanets. A larger aperture to increase sensitivity, and a smaller on sky pixel projection in which to concentrate starlight, would make this possible.
Cassini is not capable of this, but makes an ideal test facility to confirm the potential of the technique.

We have demonstrated the recovery of spectral variations in angular diameter of Omicron Ceti which have been shown to be consistent with those previously published.
The epochs studied were unfortunately too near each other, and too near to a maxima, to record any trends in Uniform Disc size with pulsation phase.

The behaviour of the instrument was analysed and modelled.
It was found that for shorter wavelengths it is possible to observe stars as dim as second magnitude in J, but for full spectral coverage it is limited to those brighter than 0 in K.
The 1\:$\sigma$ noise level for VIMS measurements was found to be consistently $<=$1\%, and the uncertainty on model fits was $<$10\%.

The Cassini data archive contains more than 100 occultation events in which a range of stellar targets were observed.
The methods described here should enable unique new astrophysical data to be produced from this unique facility, and may also motivate potential dedicated observations in the remaining years of the spacecraft's lifetime.

\section*{Acknowledgements}
The authors would like to thank Andrea Richichi for his insightful, constructive and valuable review, and Henry Woodruff for assistance with comparison to interferometric data.
This work makes use of data obtained through the Cassini-VIMS team and used resources of the rings and navigation nodes of NASA's Planetary Data System.

\bibliographystyle{mn2e} 
\bibliography{library}

\bsp

\label{lastpage}

\end{document}